# Interplay between tightly focused excitation and ballistic propagation of polariton condensates in a ZnO microcavity

R. Hahe[1], C. Brimont[1], P. Valvin[1], T. Guillet[1,*], F. Li[2,3], M. Leroux[2], J. Zuniga-Perez[2], X. Lafosse[4], G. Patriarche[4], S. Bouchoule[4]

1: Laboratoire Charles Coulomb (L2C), UMR 5221, CNRS-Université de Montpellier, Montpellier, F-FRANCE

2: CRHEA-CNRS, Rue Bernard Gregory, 06560 Valbonne, FRANCE

3: Université de Nice Sophia-Antipolis, 06103 Nice, France

4: LPN-CNRS, Route de Nozay, 91460 Marcoussis, FRANCE

* e-mail: Thierry.Guillet@umontpellier.fr

**Abstract**

The formation and propagation of a polariton condensate under tightly focused excitation is investigated in a ZnO microcavity both experimentally and theoretically. 2D near-field and far-field images of the condensate are measured under quasi-continuous non-resonant excitation. The corresponding spatial profiles are compared to a model based on the Gross-Pitaevskii equation under cylindrical geometry. This work allows to connect the experiments performed with a small excitation laser spot and the previous kinetic models of condensation in a 2D infinite microcavity, and to determine the relevant parameters of both the interaction and the relaxation between the reservoir and the condensate. Two main parameters are identified: the exciton-photon detuning through the polariton effective mass and the temperature, which determines the efficiency of the relaxation from the reservoir to the condensate.

I. Introduction

Polaritons are able to propagate over tens to hundreds of microns [1]. The analogy between the polaritons' trajectory under a constant cavity gradient and the free fall evidenced the ballistic character of this propagation [2,3]. The momentum of a polariton condensate can also be controlled either by resonant excitation (through the wavevector of the exciting laser) or under non-resonant excitation (through the spatial shape of the exciton reservoir). This last feature has been demonstrated in the case of 1D polariton condensates, showing the intricate roles of propagation and reservoir repulsion in the generation and amplification of the polariton condensate [4]. This has motivated many demonstrations of polariton devices based on condensates put into motion in an analog way to electrons in microelectronic devices; those include polariton transistors [5] and polariton tunnelling diodes [6], and proposals of optical amplifiers [7]. The propagating character of polaritons is also underlying striking features of the condensates (vortices [8,9], solitons [10–13], …), so that they are now considered as "quantum fluids" governed by non-linear hydrodynamics [14].

Most of the recent demonstrations of polariton lasing in ZnO, GaN and organic microcavities have been performed under the so-called « tightly focused excitation regime », where the excitation laser needs to be focused over a few microns spot in order to reach the threshold for polariton lasing [15–24]. This is due both to the high excitation density required for condensation and, most importantly, to the presence of photonic disorder or inhomogeneities in the cavity. As discussed by M. Wouters *et al.* [25,26], this implies that the propagation of the polariton condensate plays an important role in its formation, even though the emission is usually spatially integrated during the spectroscopy experiments.

The aim of the current work is to provide a deep insight into the interplay between the tightly focused excitation commonly employed by the room-temperature polaritonics community, and the formation and propagation of the polariton condensate and of its excitonic reservoir. This interplay will be illustrated by monitoring the spatial distribution of both the polariton condensate and the reservoir below and above threshold. The condensate distribution will then be compared to a model describing the ballistic propagation of the condensate under the repulsion of the reservoir in a cylindrical geometry and neglecting polariton relaxation. This allows evaluating whether the condensation threshold only depends on the local density of the exciton reservoir, or if it is influenced by the outwards flux of polaritons. In this last regime the condensation threshold density is increased compared to the case of an infinite 2D condensate, and this increase will be estimated. Boundary values of the corresponding physical parameters can be then extracted. The study is performed on a high Q planar microcavity with low photonic disorder [20,27] but displaying regions with either a steep photonic potential (induced by a relatively large cavity thickness gradient) or with an almost flat photonic landscape. The interplay between condensation and propagation is investigated as a function of the detuning of the polariton mode, of the temperature (at 80K and 300K), as well as of the presence or absence of a thickness gradient in the cavity.

II. The sample and its local photonic landscape

The investigated sample is a ZnO-bulk microcavity displaying polariton lasing over a wide range of exciton-photon detunings and temperatures. Earlier works described its fabrication [27] and the measured and modeled phase diagram of polariton condensation [20]. The tunability of this polariton laser results from the high crystalline quality of the ZnO active layer, the large quality factor (up to 4000) and the wide stop-bands provided by the dielectric $SiO_2/HfO_2$ Bragg reflectors.

The present work has been performed on an area where the excitons are coupled to the $4\lambda$ and $4.5\lambda$ cavity modes (comparable to Fig. 4 in Ref. [20]). This large cavity thickness assures efficient heat dissipation and prevents any undesirable heating even at high pumping intensities. Prior to the investigation of polariton propagation, it is necessary to measure the photonic landscape that is felt by polaritons. The photoluminescence spectrum (Figure 1.a) consists of multiple transitions related to each of the polariton branches (each one of them associated to the coupling of excitons to distinct cavity modes) and the bare excitons. In order to enhance the spatial resolution of the photonic potential map, confocal microscopy has been used instead of µPL experiments spatially integrating all the emission. In a confocal µPL configuration, the sample is excited with a non-resonant continuous-wave laser ($\lambda$=266 nm) far from the condensation threshold (thereby preventing any density-dependent blueshift). The excitation spot has a diameter of the order of 1 µm; it is scanned over 60x60 µm$^2$ with 1µm steps. The spatial filtering is provided horizontally by the entrance slit of the spectrometer, and vertically by selecting a single row of the CCD detector at the output of the imaging spectrometer. The confocal resolution is 0.7x1.5 µm$^2$ (as measured in ref. [28]). The confocal microscopy allows to enhance the spatial resolution of such a map, compared to a µPL experiment spatially integrating all the emission. Figure 1.b presents a mapping of the polariton mode (LPB1), arising from the coupling of ZnO excitons with the $4.5\lambda$ cavity mode. The precise modeling of the polariton dispersions is discussed into details in the Annex A. The coupling parameters are presented in table 1. At the point A, the detuning of the LPB1 branch is equal to $5 \pm 10 \ meV$. The corresponding Rabi splitting is around $280 \pm 30 \ meV$.

The investigated area presents various cavity thickness gradients depending on the exact location. For example, a small gradient (<0.3 meV.µm$^{-1}$) is measured for the branch LPB2 at the point named A, whereas a much stronger gradient (0.8 meV.µm$^{-1}$ for LPB2) is found at point B. Those two points have been chosen for the detailed propagation imaging performed in the next section. The corresponding energy gradients for the polariton branch arising from the $4.5\lambda$ cavity mode (LPB1), which displays a larger photonic fraction, worth 0.5 and 1.5 meV.µm$^{-1}$ respectively. We should notice that the energies of the polariton branches at both positions only differ by 10 meV, i.e. less than 5% of the Rabi splitting, so that their exciton-photon compositions are very similar.

III. 2D imaging of the exciton reservoir and the polariton condensate

The investigation of the spatial dynamics of the polariton condensate across its generation threshold requires a complete imaging of the exciting laser, the initially generated reservoir and the polariton emission from each emitting branch, below and above the condensation threshold. We here name "reservoir" the ensemble of all particles (excitons, polaritons with large wave-vectors) able to relax towards the condensate and emitting at energies close to the bare exciton energy. This study is performed through two-dimensional tomography. Contrary to the confocal imaging setup described in the previous section, the sample is now excited with a fixed laser spot; its emission is collected by the microscope objective and imaged by an UV achromatic lens on the entrance plane of the imaging spectrometer. The entrance slit filters the signal originating from a slice of the emission that is then spectrally dispersed by the grating and recorded by the CCD detector. The motorized translation of the last lens in the direction perpendicular to the slit allows reconstructing the full 2D image of the emission with spectral resolution. As in part of our previous works [15,20] the polariton condensate is generated under quasi-cw optical pumping with a Q-switched laser providing 400 ps pulses at 266 nm (repetition rate: 4 kHz) since polariton condensation cannot be reached under cw excitation at this wavelength.

III.1) Imaging the polariton condensation in a branch at zero detuning

Let us first investigate the condensation process at a temperature of 300 K. The power-dependent series of spectra across polariton condensation are presented in figure 2. The spectra consist of two transitions attributed to each polariton branch. Condensation is observed on the polariton branch close to zero detuning (LPB1). A blueshift of about 4 meV is measured at threshold, reflecting the repulsive potential induced by the generated excitons and felt by the polaritons. The spectra are measured with a low resolution in order to observe all transitions in a single acquisition.

The integrals of each of the transitions, as well as the weak signal corresponding to the scattered excitation laser, are then calculated at each of the points in the 2D emission plane (near field image). An additional lens allows projecting the back focal plane of the microscope objective onto the entrance plane of the spectrometer; the 2D Fourier plane of the emission is therefore measured under the exact same excitation conditions (far field image). The corresponding images ($P = 1.7\, P_{th}, T = 300K$) are presented in figure 3 (near field) and figure 4 (far field), as measured at the two points named A and B on figure 1. The main informations deduced from the spectra and the tomographies are presented in the table 2. The features are very similar at both points:

(i) Laser spot (Figure 3.a,b): it extends over 4 µm FWHM and it can be fit by a Gaussian. However, it presents tails in some specific directions, which are identical at both points. They are attributed to the multimode character of the Q-switched laser source. The available laser power being close to the polariton laser threshold, it is unfortunately not possible to perform any spatial filtering of the laser modes in order to suppress these tails before exciting the sample.

(ii) Uncondensed LPB0 and LPB1 branches: The spatial distribution of the uncondensed LPB0 branch (Fig. 3.c,d), as well as the one of the LPB1 branch below threshold (not shown) are slightly broader than the laser spot (5-6 µm FWHM). The distribution of LPB0 is centered at the laser spot in the case of point A (almost flat photonic landscape) whereas it shifts by about 3 µm in the case of point B (in the presence of a thickness gradient). The corresponding 2D far-field patterns are described in the Annex B. They present a cylindrical symmetry at point A that is broken due to the photonic gradient at point B. This reflects the impact of the photonic gradient on the polariton relaxation in a mostly photonic branch, leading to a non-zero average velocity of LPB0 polaritons. Even if the propagation properties of uncondensed polaritons are beyond the scope of the present article, those features are a clear signature of the presence of a thickness gradient.

(iii) Condensed LPB1 branch: The spatial distribution of the polariton condensate (LPB1, figure 3.e,f) presents structured patterns; contrary to the images of the exciton reservoir and the uncondensed polaritons, it is not monotonically decreasing with the distance to the center of the laser spot, and it presents sharp angular patterns. The images are very similar at both points A and B. The far field images of the condensate (Figure 4), recorded at the same points, are characterized by a minimum of the signal at $k = 0$, a first broad emission ring at $k \approx 2\ \mu m^{-1}$, and additional rings or portions of rings at larger wavevector, as indicated by the red dotted circles. They may extend beyond the accessible angular range of our present microscope objective (Numerical Aperture 0.4, i.e. $k < 6\ \mu m^{-1}$). The vanishing signal at zero wavevector and the well-defined wavevector of the condensates (see Figures 4.d and 4.f) is characteristic of a ballistic ejection of the condensate generated at the excitation spot and repelled by the generated excitons [4,16,29]. This will be modelled in detail in section IV. The absence of a proper cylindrical symmetry of the near field and far field images and the similarities observed at both sample positions lead us to conclude that the precise shape of the condensate is mostly governed by the shape of the excitation laser and its distortion compared to a purely monomode Gaussian spot. This proves that under the current experimental conditions the shape of the condensate is not reflecting the local potential felt by polaritons, induced by photonic or excitonic disorder or gradients in the investigated cavity. This is a first indication of a strong difference with the condensation in the ZnO cavity investigated in reference [30], where the disorder plays a major role in the patterning of the condensate. This allows us to compare the measured condensate profiles to a model that does not take photonic gradients into account and that does not include disorder. Let us finally emphasize that the complexity of the condensate spatial patterning is under-estimated when only cross-sections of the far-field image are recorded (Figure 4.d,f).

III.2) Imaging the polariton condensation in a branch at positive detuning

The same study has been performed at T=80K (figure 5 and table 2). The two main differences with the room-temperature case are the following:

(i) The relaxation of excitons towards the various polariton branches favors the most excitonic polaritons, so that condensation is first observed on the LPB2 branch (detuning $+200 \pm 40 \; meV$). This is consistent with the systematic study presented in [20]: the phonon-assisted relaxation is less efficient for this temperature, so that the relaxation kinetics are mainly governed by exciton-exciton scattering and, therefore, condensation is now observed in LPB2, whose excitonic fraction is larger than that of LPB1.

(ii) A transition close to the energy of the uncoupled excitons is now observed at 3.34-3.37 eV; it is attributed to higher order polariton modes, that are almost purely exciton-like [14,20], and the emission of uncoupled excitons. The 2D spatial image of this last transition provides direct access to the spatial distribution of the reservoir, which is fitted by a Gaussian profile. The diameter of the reservoir (4.5 µm FWHM) is comparable to the one of the laser spot (4 µm FWHM). The LPB1 branch (now at a slightly negative detuning of $-60 \; meV$ due to the temperature variation of the exciton energy) presents a distribution very similar to the one of the exciton reservoir (Figure 5.b).

(iii) The profile of the LPB2 emission differs from the one of the exciton reservoir and the LPB1 branch since it is much sharper near $r = 0 \mu m$ (Figure 5.b), leading to a FWHM twice smaller than the one of the exciton reservoir. The relative intensity of this sharp component compared to the long tails increases when the excitation power is increased beyond threshold. The situation is therefore different from the case of a condensate at zero detuning investigated in section III.1, where the increase of the condensate particle number led to an outward propagation of the condensate and a profile with a maximum at $r = 2 \mu m$.

IV) Model

IV.1) Generation and propagation of the polariton condensate in a cylindrical geometry

The role of propagation in the polariton condensation was theoretically explored in the seminal work of M. Wouters *et al*. [25,26]. They defined, in particular, the "tightly focused excitation regime" that corresponds to our present experimental conditions, as well as that employed in many other works on polariton condensates. Here we follow this model, assuming a cylindrical symmetry and neglecting the energy relaxation of the polaritons. This model is well adapted to polariton condensates that propagate in a ballistic way, i.e. with a well defined wavevector at a given position, as we observed in figure 4.d,f. The specificity of the present work lies in the ability to determine most of the parameters in the case of bulk-ZnO polariton condensates, or provide bounds to them, from the comparison with a detailed set of experiments.

The model describes the kinetics of the exciton reservoir and a single polariton condensate. The density $n_R$ of the reservoir is described by a rate equation

$$\frac{dn_R(r)}{dt} = P(r) - \gamma_R \, n_R(r) - R \, n_R(r) \, |\psi(r,t)|^2. \qquad (1)$$

The condensate wavefunction $\psi(r)$ is obtained in a mean-field approximation as the solution of the Gross-Pitaevskii equation (GPE) in the absence of any external potential:

$$i\hbar \frac{\partial \psi(r,t)}{\partial t} = \left( \hbar\omega_0 - \frac{\hbar^2}{2\,m^*}\nabla_r^2 + \frac{i\hbar}{2}\bigl(R\,n_R(r) - \gamma_{pol}\bigr) + \hbar g_R \, n_R(r) + \hbar g \, |\psi(r,t)|^2 \right) \psi(r,t), \quad (2)$$

where $\hbar\omega_0$, $m^*$ and $\gamma_{pol}$ are respectively the energy, the effective mass and the decay rate of the investigated polariton branch; $g_R$ and $g$ are the exciton-polariton and polariton-polariton interaction constants. The stimulated relaxation from the reservoir to the condensate is accounted for through the term $R\,n_R(r)$, depending linearly on the reservoir density. The reservoir consists both of excitons, with wavevectors beyond the light cone in the ZnO active layer, and high energy polaritons beyond the so-called bottleneck region; its decay rate is denoted $\gamma_R$. A Gaussian profile is chosen for the pumping rate in the reservoir, $P(r)$, according to the measured exciton reservoir (4.5 µm FWHM). We neglect here the terms corresponding to the disorder and/or the photonic gradient in the microcavity, as discussed at the end of the section III.1.

Following the approach developed for small excitation spots under stationary excitation and cylindrical symmetry in ref. [26], the condensate wavefunction at a given blueshift ($\hbar\omega_c - \hbar\omega_0$) writes $\psi(r,\theta,t) = \psi_m(r)e^{-i\omega_c t}e^{im\theta}$. In this work we only consider the vortex-free case of $m = 0$ (no angular momentum in the condensate). For radii $r$ much larger than the spot size, the stationary solution freely propagates with a wavevector $k_c = \sqrt{2m^*/\hbar(\omega_c - \omega_0)}$ and vanishes due to the finite polariton lifetime, so that it asymptotically follows the Hankel function $H_0^{(1)}\left(\sqrt{2m^*/\hbar\left(\omega_c - \omega_0 + i\gamma_{pol}/2\right)}.r\right)$. For $m = 0$ (no vortex) and an experimentally determined blueshift $\hbar(\omega_c - \omega_0)$, the full condensate wavefunction $\psi_0(r)$ is numerically calculated with a 4$^{th}$ order Runge-Kutta algorithm.

The numerical resolution of the problem is then performed both in the case of a non-depleted reservoir ($n_R(r) = P(r)/\gamma_R$) and in the case of a depleted reservoir ($n_R(r) = P(r)/(\gamma_R + R|\psi_0(r)|^2)$), as deduced from eq. (1). The two assumptions will be compared in section V in order to conclude about the role of depletion in the condensation dynamics.

### IV.2) A procedure for the choice of the model parameters

The determination of the parameters of the Gross-Pitaevskii equation has a strong influence on the obtained solution. Some parameters have been directly extracted from measurements: time-resolved photoluminescence experiments give access to the exciton reservoir lifetime $\tau_R = 40\ ps$, leading to a reservoir recombination rate $\gamma_R = 0.016\ meV$. The cavity decay rate $\gamma_{cav} = 0.8\ meV$ is deduced from the measured quality factor $Q = 4000$ [27]. Each polariton branch is characterized by an effective mass that is measured in far-field dispersion for LPB0 and LPB1 (at T=300K), and deduced from transfer-matrix simulations for the heaviest branch LPB2 (observed only at T=80K). The full set of parameters for the polariton branches is presented in table 3.

Three parameters ($g_R, g$ and $R$) are unknown from experiments. The polariton interaction constants $g_R$ and $g$ are assumed to depend on the Hopfield coefficient of the investigated LPB, and the exciton-exciton interaction constant $g_{XX}$. As discussed in the Annex A, we prefer here to introduce the coefficient $x = \frac{\partial E_{LPB}}{\partial E_X}$ instead of the Hopfield coefficient, so that the polariton-reservoir and polariton-polariton interaction constants read $g_R = x \cdot g_{XX}$ and $g = x^2 \cdot g_{XX}$. The parameters $g_R$ and $g$ have been strongly debated in the study of GaAs microcavity polaritons, and are still unknown for ZnO microcavities. Theoretical predictions in the case of interacting 3D excitons in a slab [31] (corresponding to our bulk ZnO microcavity) lead to a value $\hbar g_{XX}^a \approx 10\ E_b\ a_B^3/L \approx 1.8\ 10^{-6}\ meV.\mu m^2$, where $E_b = 60\ meV$ and $a_B = 1.4\ nm$ are the ZnO binding energy and Bohr radius; $L = 890\ nm$ is the thickness of the ZnO active layer at the investigated point as discussed in section II.

A second line of reasoning can be followed in order to determine those parameters: the parameter $g_R$ can be also accessed through the measured blue-shift of the polariton line at threshold, $\hbar(\omega_c - \omega_0) = \hbar g_R \cdot n_{R\ th}$. Even if there is no independent experimental determination of the exciton density at threshold $n_{R\ th}$, it has been calculated within a rate equation model in the 2D case of an infinite spot size in the same microcavity [20,32]: $n_{R\ th}^{2D} \approx 5.10^4 \mu m^{-2}$ at room temperature, which can be imposed to this value in our simulations. Following this approach, a second value of the interaction parameter will be deduced from the simulations shown in this section: $\hbar g_{XX}^b \approx 1.0\ 10^{-5}\ meV.\mu m^2$, that is of the order of 6 times larger than $g_{XX}^a$. This apparent discrepancy will be discussed in section V.

The gain rate $R$ is phenomenological; in the 2D case of an infinite spot size, it is related to the reservoir density at threshold and the polariton decay rate $\gamma_{pol}$ through $R\ n_{R\ th}^{2D} = \gamma_{pol}$ since gain and losses exactly compensate at the laser threshold. Contrary to the polariton interaction parameters, it should depend on the temperature and the detuning of the polariton branch since it reflects the efficiency of the stimulated relaxation from the exciton reservoir to the polariton condensate. Again its determination relies on the knowledge of $n_{R\ th}^{2D}$.

In order to easily compare with the results of the rate equation model presented in [20], we have chosen the set of parameters based on $n_{R\ th}^{2D} \approx 5.10^4 \mu m^{-2}$, i.e. $\hbar g_{XX}^b \approx 1.0\ 10^{-5}\ meV.\mu m^2$ and $\hbar R = 5.10^{-6}\ meV.\mu m^2$ at $T = 300K$. Since the density of the exciton reservoir only appears through the terms $R\ n_R$ and $g_R\ n_R$ in the master equations (1) and (2), it should be noticed that in the absence of any strong reservoir depletion or strong polariton-polariton interactions, as we will show, the model leads to identical results for the condensate if we use the other set of parameters, $\hbar g_{XX}^a \approx$ $1.8\ 10^{-6}\ meV.\mu m^2$ and $\hbar R = 8.6\ 10^{-7}\ meV.\mu m^2$, and exciton densities 6 times larger in the reservoir (see table 3). This point will be further discussed in section V.

### IV.3) Simulations of a polariton condensate at zero detuning

The simulations corresponding to the experimental results of figure 3 are presented on figure 6. The adjustment of the simulation parameters to the experiment is performed in 3 steps:

(i) The known blueshift (4 meV at $P_{th}$, 12 meV at 1.7 $P_{th}$) and the effective mass of the polariton branch LPB2 determine the value $k_c$ of the polariton wavevector far from the reservoir.

(ii) The long-distance tails of the condensate profile are compared to the model, for radii larger than the reservoir FWHM. The slope of the profile tails requires a slight adjustment of the polariton decay rate $\gamma_{pol}$.

(iii) Finally the density of the exciton reservoir is adjusted to reproduce the condensate pattern close to the center of the spot.

In figure 6.a, the polariton decay rate is taken to $\gamma_{pol} = 0.35\ meV$, i.e. almost half of the photon decay rate $\gamma_{cav}$ obtained from earlier linewidth measurements. This corresponds to a polariton lifetime twice longer than the cavity lifetime. This factor 1/2 is consistent with the 1/2 photonic fraction of the LPB1 polariton branch. When the pumping rate is increased ($P = 1.7\ P_{th}$, figure 6.c), the condensate profile (for distances smaller than 10 µm) decreases more rapidly with r than for $P = P_{th}$, despite a larger wavevector as deduced from the larger blueshift. This can be accounted for with a larger polariton decay rate $\gamma_{pol} = 1.8\ meV$ due, for example, to the presence of additional decay channels such as scatterings towards other polariton states. This will be discussed in section V.

The determination of the density of the exciton reservoir is more straightforward. The total number of excitons in the reservoir is obtained to be $N_R = 1.6\ 10^7$ (resp. $4.9\ 10^7$) for $P = P_{th}$ (resp. $P = 1.7\ P_{th}$), corresponding to an exciton density at the center of the spot $n_R(r = 0) = 8\ 10^5\ \mu m^{-2}$ (resp. $2.4\ 10^6\ \mu m^{-2}$). The increase of $n_R$ by a factor 3 instead of 1.7 shows that the assumption of a constant relaxation efficiency from the laser energy to the reservoir and/or from the reservoir to the condensate does not fully correspond to the experimental situation.

The formation of a condensate pattern with a local minimum at the laser spot center and local maxima at $r = 2\ \mu m$ is a striking feature of the polariton near-field image presented in figure 3.e,f. Even if the experimental pattern is not presenting a cylindrical symmetry as assumed in our model, it is qualitatively reproduced in the simulations. It should be noticed that for $P = 1.7\ P_{th}$ the pattern profile is very sensitive to the reservoir density, so that any inhomogeneity of the excitation or of the relaxation efficiency induces large variation of the condensate local density due to the non-linearity of the formation process; such inhomogeneities have a weaker effect for $P \leq P_{th}$.

The analysis of the local contributions to the variation of the condensate distribution allows a better understanding of the condensate formation mechanism. As discussed in ref. [26], the polariton conservation equation in the stationary regime is the sum of three terms (gain, polariton decay and polariton current):

$$(R\ n_R(\boldsymbol{r}) - \gamma_{pol})|\psi(\boldsymbol{r})|^2 - \frac{\hbar}{m^*}\ div(|\psi(\boldsymbol{r})|^2.\boldsymbol{k}(\boldsymbol{r})) = 0, \qquad (3)$$

where the local polariton wavector $k(r)$ is obtained as the gradient of the phase of the polariton wavefunction. At the condensation threshold ($P = P_{th}$, figure 6.b), even if the condensate profile has its maximum at $r = 0$, the polariton current is much larger than the polariton losses, so that the reservoir density $n_R(0) = 8\ 10^5\ \mu m^{-2}$ is 16 times larger than $n_{R\,th}^{2D}$ in the 2D case. At $P = 1.7\ P_{th}$ the polariton current and the polariton decay contributions have comparable magnitudes. The polariton condensate radially accelerates and gets amplified on the sides of the exciton reservoir, with a maximum at $r = 2\ \mu m$; at this position, the stimulated relaxation from the exciton reservoir therefore feeds the condensate with polaritons with a non-zero wavevector.

The analysis of the k-space distribution of the condensate is further illustrated in figure 7. In order to compare with the experimental cross-section of the far-field pattern of the condensate (figure 7.a), the local polariton wavector $k(r)$ is plotted as a white line on figure 7.b, and the emission intensity at the same position is represented in false colors. Its spatial average provides the simulated far-field pattern of the condensate (figure 7.a). The simulated and experimental profiles both present a maximum at $k \neq 0$, a signature of the ballistic propagation of the polaritons. However the wavevector of this maximum is very different: $k = 2\ \mu m^{-1}$ for the experiment and $5\ \mu m^{-1}$ for the simulation. Much weaker peaks are observed at $4 - 5\ \mu m^{-1}$ in the experimental profile, corresponding to partial rings in the 2D far-field pattern (dotted circles in figure 4.c,e). This will be discussed in the section V.

IV.4) Simulations of a polariton condensate at positive detuning

The formation process of the condensate strongly depends on the excitonic character of the polariton branch, i.e. its exciton Hopfield coefficient and more importantly its effective mass, as shown in the following analysis of the experiment performed on LPB2 at $T = 80K$ (Figure 5). Here the effective mass of the polariton branch is almost 5 times larger than the one of the LPB1. The experimental profile of the condensate ($P = 1.3\ P_{th}$, figure 8.a) is dominated by a sharp maximum at $r = 0\ \mu m$, and propagation tails. The measured blueshift is 11 meV, leading to an asymptotic polariton wavevector far from the spot center of $9\ \mu m^{-1}$ (beyond the numerical aperture of our microscope objective) and an exciton density $n_R(r = 0) = 1.2\ 10^6\ \mu m^{-2}$, 20 times larger than the calculated threshold density for an infinite 2D condensate. The parameters of the corresponding simulation are $\gamma_{pol} = 0.35\ meV$, as in figure 6.a, and $R = 4.9\ 10^{-7}\ meV.\mu m^{-2}$ (lower than at T=300 K). Figure 8.b shows that the condensate is mainly generated at the center of the spot, and then propagates outwards without any amplification, so that the condensate profile is close to the Hankel function.

V. Discussion

The quantitative analysis of the spatial distribution of the polariton condensate in the investigated ZnO microcavity shows that different formation schemes can be involved depending on the detuning of the polariton branch and the excitation density. The most important parameter is the effective mass of the investigated polariton branch. For a given blueshift $\hbar(\omega_c - \omega_0)$, it determines the propagation speed of the polariton wavepacket far from the reservoir $v_c = \sqrt{2\ \hbar(\omega_c - \omega_0)/m^*}$, and the order of magnitude of the time spent by the polaritons within the reservoir before free propagation, $t_1 = \sigma_R/v_c$ (0.24ps for LPB1 vs 0.5 ps for LPB2 according to our simulations; $\sigma_R$ is the waist size of the Gaussian profile of the reservoir). This time $t_1$ has to be compared to the timescale of the stimulated relaxation from the reservoir to the condensate, $t_2 = 1/(R\ n_R)$, that is estimated to 0.07ps for LPB1 at 300K, vs 0.8ps for LPB2 at 80K. In the case of a "light" polariton branch (LPB1) and an efficient relaxation (T=300K), the stimulated relaxation is faster than the escape time out of the reservoir, so that the polariton condensate gets amplified as it flows away. This situation presents some analogy with the amplification of a polariton condensate after propagation and reflection, demonstrated in 1D polariton ridges [4]. In the case of a "heavy" polariton branch (LPB2) and a less efficient relaxation (T=80K), the condensate forms at the center of the reservoir spot, and then freely propagates outwards without amplification, like in a "free-fall". The difference between those two regimes can be evidenced through the complementary measurements of the real-space and k-space distributions of the polariton condensate. They could not be distinguished in previous studies based only on far-field measurement [16]. Due to the strong correlation between the relaxation efficiency and the detuning of the condensed branch in our multi-mode ZnO microcavity, two scenario cannot be explored experimentally in our system, corresponding to an efficient relaxation to a condensate in a heavy LPB, and an unefficient relaxation to a condensate in a light LPB.

The "tightly focused excitation regime" leads to an increase of the threshold reservoir density for condensation, compared to the 2D case of an infinite excitation spot. As discussed in section IV.2 and illustrated in the two parameter sets in table 3, the estimate of the exciton density at threshold is strongly dependent on the choice of the parameters for interactions ($g_R$) and stimulated relaxation (R). It also relies on a broad set of experimental results, so that its precise determination is challenging. However our simulations provide the first estimates of this increase for the investigated case of a non-resonant quasi-cw excitation (400 ps pulses at 4.66 eV, spot diameter 4 μm FWHM). The threshold ratio $n_{R\,th}(0)/n_{R\,th}^{2D}$ is of the order of 10 to 20 times, with an uncertainty estimated to a factor 2.

The absolute value of the threshold density for condensation cannot be exactly determined. We consider here only one reservoir of excitons, that is involved both in the condensate repulsion (coefficient $g_R$) and in the stimulated relaxation forming the condensate (coefficient $R$), leading to a discrepancy with previous theoretical predictions of about one order of magnitude. This may be related to a more complex situation where all photo-generated carriers contribute to the condensate repulsion, whereas only a fraction of them populate the reservoir of excitons and large-k polaritons that can efficiently feed the condensate. Assuming the existence of two distinct reservoirs is a possible way to explain this finding: the exciton reservoir, composed of excitons and large-k polaritons, is involved in the stimulated relaxation term $\hbar R\ n_{R1}$ of the equation (2), whereas a second reservoir composed of all the photo-generated carriers (all electron-hole pairs, some of them not yet relaxed to the exciton energy or not able to undergo stimulated relaxation towards the condensate) contribute to the repulsive potential $\hbar g_R\ n_{R2}$ felt by the condensate. This "two reservoirs" assumption is summarized in the table 3, and is also debated in the formation of polariton condensates in GaAs and GaN microcavities [33–35].

A second limitation of the present model lies in the absence of relaxation, that was recently taken into account in the theoretical modeling of polariton condensation [36]. Indeed we don't observe energy relaxation of the polaritons within the condensed polariton branch beyond threshold in the recorded spectra, so that the assumption of a negligible energy relaxation is consistent with our experiments. However a discrepancy is observed between the experimental k-space distribution of the polariton condensate and the simulated one; the polaritons don't reach the predicted maximum value of their wavevector far from the excitation spot. Two possible explanations can be proposed that would also require further investigations: (i) at large polariton densities in the condensate, part of the polaritons are ejected from the condensate towards other polariton branches [18,37], leading to a decrease of the polariton lifetime; our simulations suggest that even a fourfold increase of the polariton decay rate $\gamma_{pol}$ is not enough to accound for the k-space experimental results. (ii) Even if they don't relax in energy, the polaritons within the condensate may relax their wavevector due to polariton-polariton scattering.

Our model allows to estimate the populations of the exciton reservoir (absolute particle numbers of the order of $10^7$ with our set of parameters) and the condensate (of the order of $10^3$), so that we can conclude that (i) there is no depletion of the reservoir and (ii) the polariton-polariton repulsion within the condensate is negligible compared to the one of the reservoir. We should notice that the absence of depletion may be specific to the 400 ps pulsed excitation used in the present work, as well as many previous studies on GaN and ZnO polariton condensation; such pulses are long compared to the typical timescales of the excitons and photons in the system, but they are probably not long enough to reach a stationary regime where the condensate particle number is limited by the depletion of the reservoir.

## VI. Conclusions

We have studied the formation and propagation of a polariton condensate in a ZnO microcavity in the so-called "tightly-focused excitation regime". The 2D imagery of the spectrally-resolved emission in real and reciprocal spaces provides a complete set of experimental results that can be compared to a simple model based on the GPE. The respective roles of the condensate formation, the repulsion by the exciton reservoir, the condensate amplification and the condensate propagation are identified. The validity of this model is discussed in depth, as well as the possible sets of physical parameters compatible with the experiments. Two regimes are evidenced depending on the detuning of the condensing polariton branch (through its effective mass) and on the temperature (which plays a central role in enhancing polariton relaxation): light polaritons near zero exciton-photon detuning propagate slowly enough under the exciton reservoir so that their condensate is strongly amplified "on the fly", whereas heavy polaritons at positive detuning accelerate to larger wavevectors (because of a larger exciton-exciton repulsion) and are not amplified along propagation. Finally we can estimate the increase of the exciton density at threshold when comparing tightly focused excitation and the ideal case of an infinite 2D system; this factor reaches 10 to 20 in the present work. These considerations are crucial in order to properly design and predict future polariton laser devices based on micron-sized exciton reservoirs.

## Acknowledgement

The authors would like to thank F. Réveret, P. Disseix, J. Leymarie, D. Solnyshkov and G. Malpuech for fruitful discussions, J.R. Huntzinger for the transfer matrix simulations and H. Gargoubi for her careful reading of the manuscript.

**Annex A: Modeling of the polariton dispersion and the polariton-exciton interaction**

The multi-mode character of the investigated microcavity and the very large value of its Rabi splitting requires a detailed modeling of the polariton eigenstates in each lower polariton branch, in order to extract the relevant parameters for the model developed in section IV: the effective masses and the polariton-exciton interaction constants $g_R$ for each branch. $g_R$ is reduced compared to the exciton-exciton interaction constant $g_{XX}$ because the polariton is only partially excitonic in nature. Assuming that the exciton and polariton densities vary on length-scales comparable to the optical wavelength, and much larger than the exciton Bohr radius, $g_R$ is obtained by slightly shifting the exciton energy $E_X$ and deducing the corresponding variation of the polariton energy $E_{LPB}$: it reads $g_R = \frac{\partial E_{LPB}}{\partial E_X}.g_{XX}$ . Indeed in the presence of an exciton reservoir with a density $n_R$, the potential energy of an additional exciton reads $\hbar g_{XX} n_R$. The potential energy of an additional polariton can be obtained by calculating the corresponding eigenmodes of the exciton-photon system, with an exciton energy shifted by $+\hbar g_{XX} n_R$, leading to the blueshift $\hbar g_R \, n_R$ of the polariton branch in the presence of the reservoir.

The usual modeling of the polariton eigenstates is performed through a coupled oscillator model: the exciton-photon interaction is described within a 2x2 matrix, with 2 main parameters (the exciton-photon detuning and the Rabi splitting). This model allows to determine the exciton Hopfield coefficient $c_X$ of the polariton wavefunction and the corresponding exciton content $|c_X|^2$. The polariton effective mass then reads $m_{LPB}^* = m_{cav}^*/(1 - |c_X|^2)$, where $m_{cav}^*$ is the effective mass of the bare cavity mode. The interaction constant of a polariton with an exciton is $g_R = |c_X|^2 \, g_{XX}$. This model can be extended to the case of multiple cavity modes. In our present microcavity, we can include the 3 cavity modes $cav_0$, $cav_1$, $cav_2$, corresponding to the $5\lambda$, $4.5\lambda$, $4\lambda$ resonances of the active layer, leading to a 4x4 matrix.

However, due to the strength of the exciton-photon coupling, the coupled-oscillator approach is not valid for the polariton branches at positive detuning [38]. We therefore prefer to extract directly those parameters from the transfer-matrix simulations, which provide an accurate description of the polariton energy dispersions. The figure A1 presents such a comparison of transfer-matrix simulations and the corresponding coupled oscillator modeling. The agreement of the two approaches is good for most of the polariton branches, but a clear discrepancy is observed when the photon mode $cav_1$ reaches positive detunings, for angles larger than $25 \, deg$ [38]. Moreover the LPB2 branch is visible at 3.315 eV in the transfer matrix simulation, whereas it cannot be obtained in the coupled oscillator model. This comparison shows that only the transfer-matrix simulations provide a proper description of the energies of all polariton branches, including at positive detuning, because they take into account the non-perturbative character of the strong exciton-photon interaction in ZnO microcavities. The same simulations also provide the dependence of the polariton branches as a function of the cavity thickness, as shown in the figure 1 of ref. [20].

The variation of the energies of the polariton branches has been calculated for small variations of the exciton energy $x = \frac{\partial E_{LPB}}{\partial E_X}$ or the cavity thickness $\frac{\partial E_{LPB}}{\partial L}$ . The polariton-exciton interaction constant is then taken as $g_R = x \, g_{XX}$. The results are presented in the table 1, together with the detunings, effective masses, Rabi splittings of the investigated polariton branches. The Rabi splitting of the investigated 4.5 $\lambda$ cavity mode is slightly larger than the one measured for a 2.5 $\lambda$ cavity mode in ref [20]. The estimate of the parameter $x$ is comparable to the exciton content $|c_X|^2$ at negative detuning and becomes larger at zero and positive detunings. This result may be counter-intuitive since the sum of the coefficients $x$ for all branches is larger than unity, contrary to the sum of their $|c_X|^2$ Hopfield coefficients deduced from the diagonalization of the coupled oscillator Hamiltonian. This is due to the multi-mode character of the microcavity and to the large sensitivity of the most excitonic polariton branches on the bare exciton energy.

**Annex B: Propagation of the polaritons in the uncondensed branch LPB0**

The impact of the cavity thickness gradient on the propagation of the polaritons is evidenced in the near-field emission patterns shown in figure 3. The gradient at point B induces a 3 µm shift of the uncondensed polaritons in the photonic LPB0 branch. This translation of the LPB0 polaritons can also be seen in the far-field patterns, as shown on the figure A2. At point A, the k-space distribution of the LPB0 polaritons is rather isotropic, with a large amount of emission near and beyond $4-6\ \mu m^{-1}$, i.e. the accessible numerical aperture of our microscope objective; this reflects their out-of-equilibrium distribution and the so-called relaxation bottleneck. At point B the distribution is not isotropic, with a stronger emission in one half of the observable k-space that points in the direction of the photonic gradient seen in figure 1 b; this reflects the drift of the uncondensed polaritons along the photonic gradient.

**Tables**

|  |  |  | T=300K | | | T=80K | | |
|---|---|---|---|---|---|---|---|---|
|  |  |  | LPB0 | LPB1 | LPB2 | LPB0 | LPB1 | LPB2 |
|  | Exciton $X_A$ | Energy (eV) | 3.300 | | | 3.368 | | |
| Transfer-matrix | Bare cavity modes | Energy (eV) | 3.047 | 3.305 | 3.568 | 3.047 | 3.305 | 3.568 |
|  |  | Effective mass | 3.7E-5 | 4.7E-5 | 5.0E-5 | 3.7E-5 | 4.7E-5 | 5.0E-5 |
|  |  | Detuning (meV) | -253 | 5 | 268 | -321 | -63 | 200 |
|  | Polariton branches | Energy (eV) | 3.005 | 3.181 | 3.256 | 3.012 | 3.210 | 3.315 |
|  |  | Effective mass | 3.9E-5 | 7.5E-5 | 2.9E-4 | 4.1E-5 | 8.7E-5 | 3.5E-4 |
|  |  | $x = \frac{\partial E_{LPB}}{\partial E_X}$ | 0.12 | 0.48 | 0.87 | 0.09 | 0.36 | 0.81 |
|  |  | $\frac{\partial E_{LPB}}{\partial L}$ $(meV.nm^{-1})$ | -1.7 | -1.3 | -0.34 | -1.8 | -1.7 | -0.5 |
| Coupled oscillators | Polariton branches | Energy (eV) | 3.005 | 3.181 | N/A | 3.012 | 3.210 | N/A |
|  |  | Effective mass | 4.5E-5 | 6.5E-5 | N/A | 4.5E-5 | 6.5E-5 | N/A |
|  |  | Rabi energy (meV) | 280±30 | | N/A | 280±30 | | N/A |
|  |  | $|c_X|^2$ | 0.15 | 0.32 | N/A | 0.10 | 0.27 | N/A |
| Experiment (Pt. A) | Polariton branches | Energy (eV) | 3.035 | 3.187 |  | 3.056 | 3.222 | 3.310 |
|  |  | Effective mass | 3.5E-5 | 7.1E-5 |  |  |  |  |
|  |  | Condensation |  | X |  |  |  | X |

**Table 1:** Main polariton parameters relevant for the experiments and simulations presented in the sections III and IV, as obtained from the transfer-matrix simulations at normal incidence, the coupled oscillator model, and the experiments. The cavity thickness is $L = 890\ nm$.

|  | T = 300K | | | | T = 80K | | | | |
|---|---|---|---|---|---|---|---|---|---|
|  | Laser | Reservoir | LPB0 | LPB1 | Laser | Reservoir | LPB0 | LPB1 | LPB2 |
| Condensate |  |  |  | X |  |  |  |  | X |
| Spatial FWHM | 4 µm |  | 5µm | 6 µm | 4 µm | 4.5 µm | 4 µm | 4 µm | 2 µm |
| Blueshift at threshold | $P_{th}$ = 0.94nJ per pulse |  | 3 meV | 4 meV | $P_{th}$ = 0.36nJ per pulse |  | 2 meV | 3 meV | 8 meV |
| Blueshift beyond threshold | P = 1.7 $P_{th}$ |  | 5 meV | 12 meV | P = 1.3 $P_{th}$ |  | 3 meV | 5 meV | 11 meV |

**Table 2:** Parameters of the imaging experiments under tightly-focused excitation (From figures 2, 3, 5).

|  | T = 300K | | T = 80K |
|---|---|---|---|
|  | P = P$_{th}$ | P = 1.7 P$_{th}$ | P = 1.3 P$_{th}$ |
| Condensate polariton branch | LPB1 | | LPB2 |
| Blueshift (meV) | *4* | *12* | *11* |
| $x = \dfrac{\partial E_{LPB}}{\partial E_X}$ | *0.47* | | *0.87* |
| Polariton decay rate $\hbar\gamma_{pol}$ (meV) | 0.35 | 1.8 | 0.35 |
| Exciton-exciton interaction $\hbar g_{XX}^b$ (eV.µm²) | *1.0E-8* | | |
| Exciton-polariton interaction $\hbar g_R$ (eV.µm²) | 5.0E-9 | | 9.0E-9 |
| Stimulated relaxation rate $\hbar R$ (eV.µm²) | 3.3E-9 | | 4.9E-10 |
| Exciton reservoir density $n_R(r=0)$ (µm⁻²) | 8.0E+5 | 2.4E+6 | 1.2E+6 |
| Equivalent threshold for 2D condensation (µm⁻²) | *5.0E+4* | | 5.2E+4 |
| Hankel wavevector (µm⁻¹) | 2.8 | 4.9 | 9.2 |
| Two reservoirs' assumption |  |  |  |
| Exciton-exciton interaction $\hbar g_{XX}^a$ (eV.µm²) | *1.80E-9* | | |
| Exciton-polariton interaction $\hbar g_R$ (eV.µm²) | 8.6E-10 | | 1.6E-9 |
| e-h pair reservoir density $n_R(r=0)$ (µm⁻²) | 4.6E+6 | 1.4E+7 | 6.9E+7 |
| Equivalent threshold for 2D condensation (µm⁻²) | 2.9E+5 | | 3.1E+5 |

**Table 3:** Parameters of the numerical resolution of the Gross-Pitaevskii equation, corresponding to the figures 6 to 8. The input parameters of the model are indicated in italic.

**Figures**

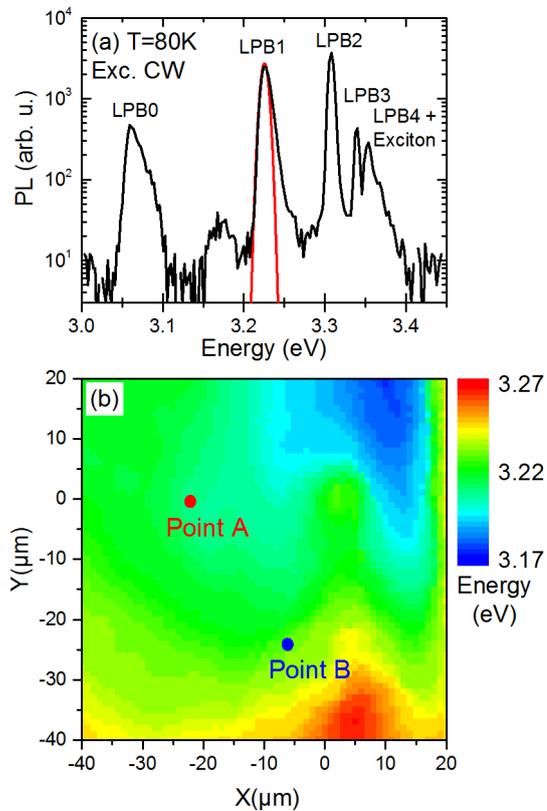

**Figure 1:** (a) Confocal micro-photoluminescence spectrum of the microcavity at T=80K, under CW excitation; the low energy part of the LPB1 line is fit with a Gaussian lineshape (red line); (b) Energy map of the LPB1 line measured under scanning confocal µPL (deduced from the Gaussian fit) at 80K; the spectrum (a) is measured at the position (-6µm,-26µm). The two points of interest investigated in the section III are labelled A and B.

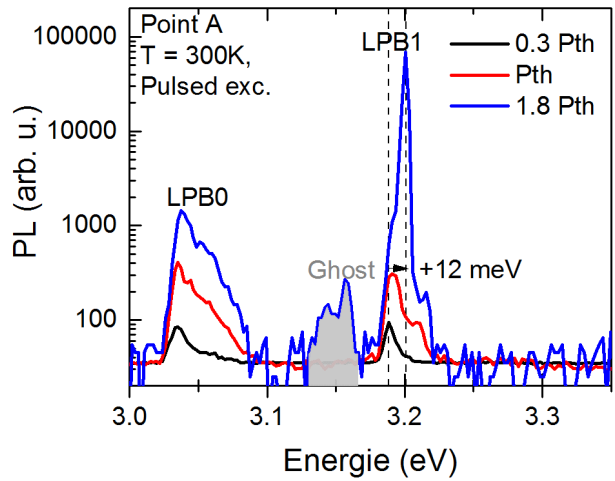

**Figure 2:** µPL spectra at point A as a function of the excitation power, under pulsed excitation at T=300K. The threshold for polariton condensation is $Pth = 0.94\ nJ/pulse$. The ghost of the spectrometer is related to the very intense peak of the polariton condensate, and indicated by a grey shade.

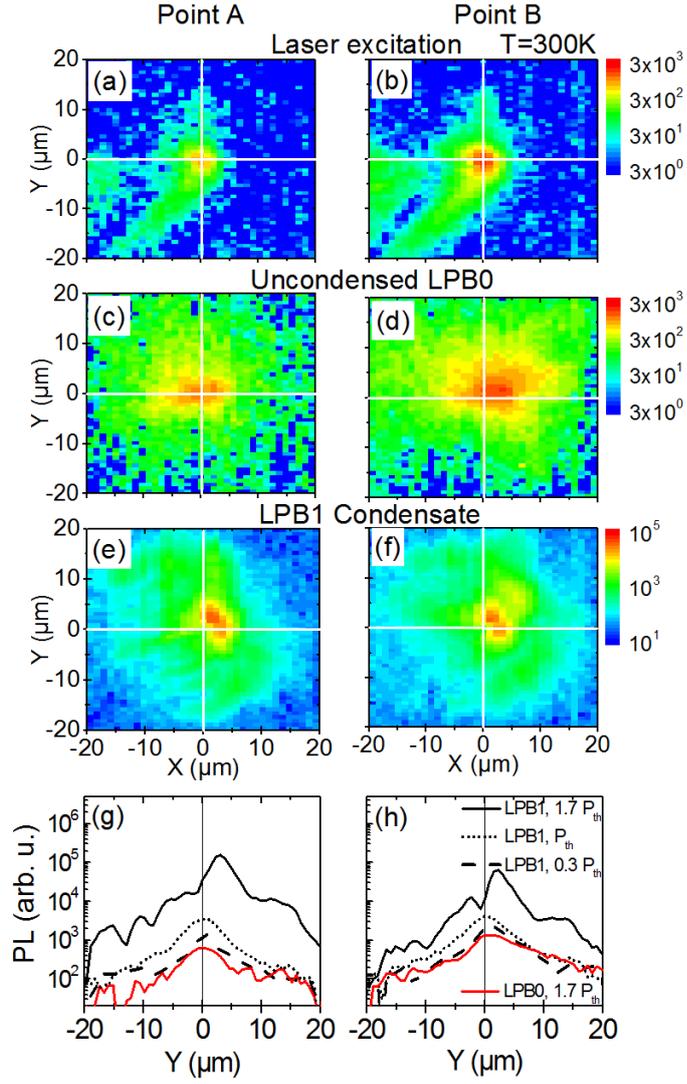

**Figure 3:** Near-field images under pulsed excitation at $P = 1.7\ P_{th}$, $T = 300K$. The points A (a,c,e,g, flat landscape) and B (b,d,f,h, slope along Y) correspond to the positions indicated on figure 1. The signal is integrated at the energy of the exciting laser (a,b), the LPB0 line (c,d) and the LPB1 line (e,f). Cross-sections are extracted along the Y direction (g,h). The FWHM of the corresponding profiles at $P_{th}$ are given in table 2.

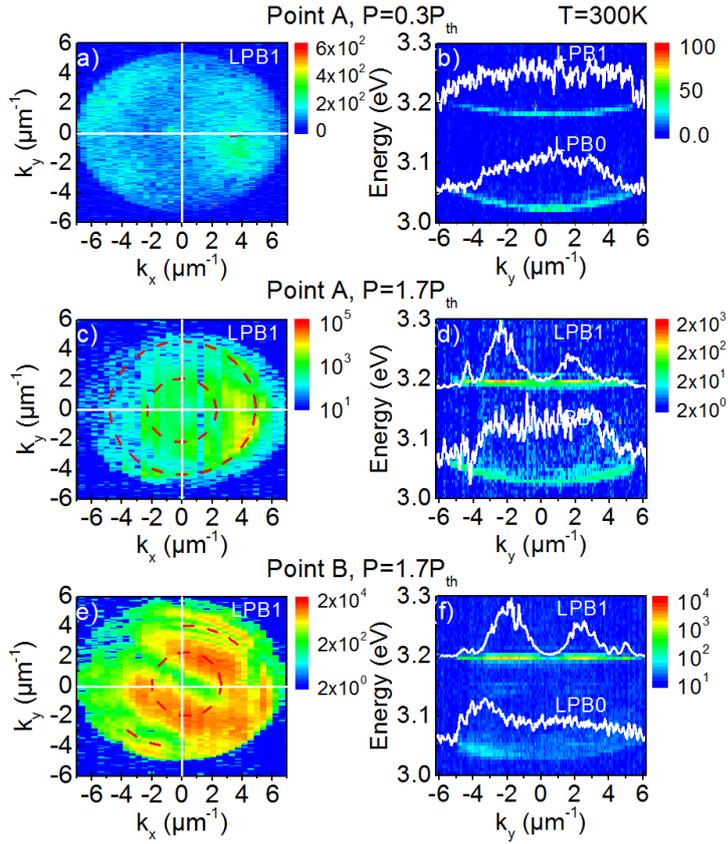

**Figure 4:** Far-field images of the LPB1 line under pulsed excitation at point A below (a) and above (c) threshold, and point B above threshold (e), $T = 300K$. The red dashed circles are guides for the eye. The corresponding spectrally-resolved vertical cross-sections extracted at $k_x = 0$ (b,d,f). Linear false color scales are used below threshold (a,b) and logarithmic ones above threshold (c-f).

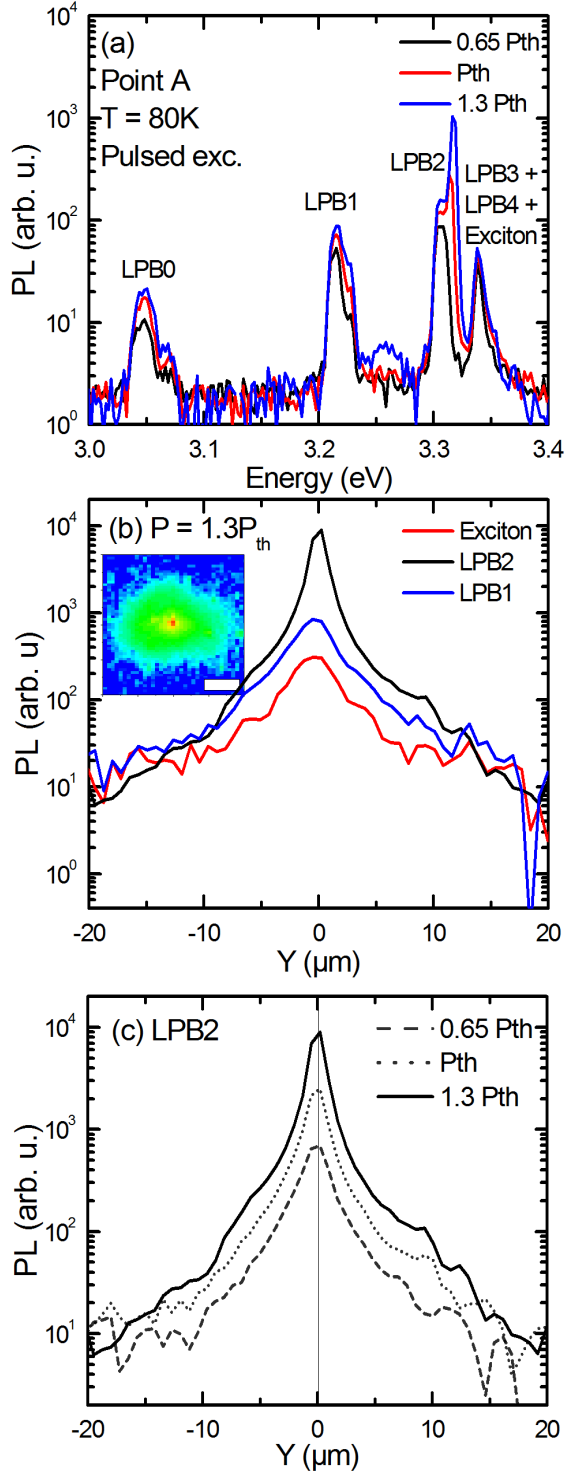

**Figure 5:** (a) μPL spectra at point A as a function of the excitation power, under pulsed excitation at T=80K. (b) Cross-sections of the exciton reservoir, the uncondensed LPB1 branch and the polariton condensate (LPB2), at $P = 1.3\ P_{th}$. The inset presents the 2D image (10μm scale bar) of the polariton condensate in false colors (logarithmic scale from blue to red). (c) Cross-section of the LPB2 emission as a function of the excitation power. The FWHM of the corresponding profiles at $P_{th}$ are given in table 2.

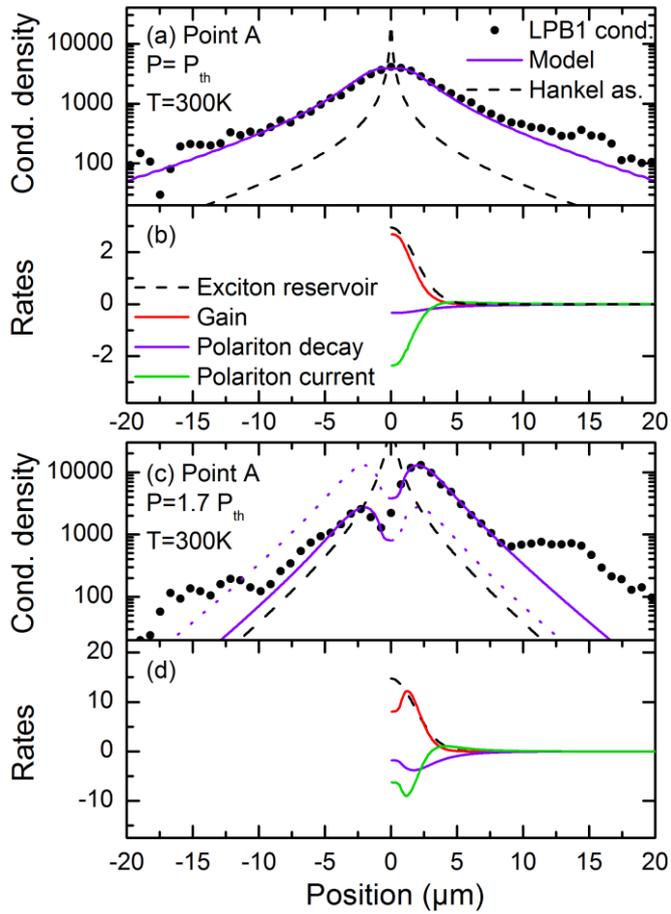

**Figure 6:** (a) Experimental y-axis profiles of the condensate (dots) at the polariton laser threshold ($P = P_{th}$, LPB1, T=300K, figure 3.f), GPE simulation (plain line) and asymptotic Hankel function (dashed line). (b) Radial dependence of the simulated rates for gain (stimulated relaxation towards the condensate), polariton decay (losses) and in-plane polariton current. (c,d) Same informations for $P = 1.7\ P_{th}$. The simulation parameters are summarized in table 3.

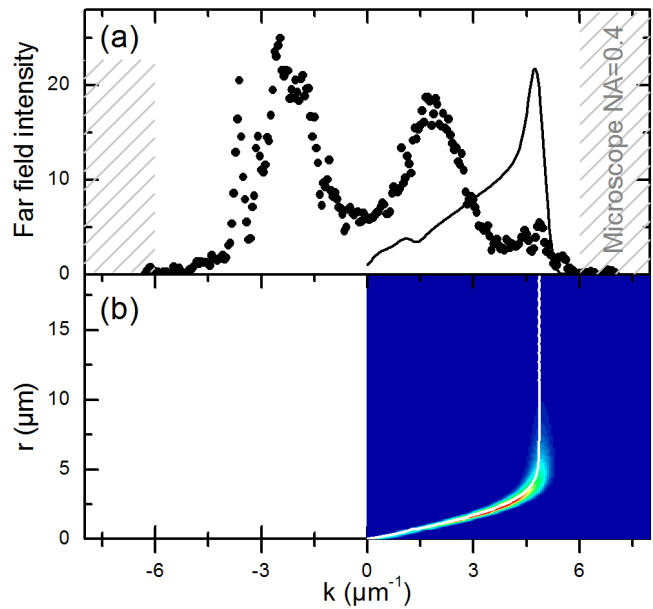

**Figure 7:** (a) Experimental $k_y$ profiles of the condensate far-field emission (dots) at the polariton laser threshold ($P = 1.7\ P_{th}$, LPB1, T=300K), and the GPE simulation corresponding to figure 6.c,d (plain line) (b) Relation between wavevector and radius along the condensate propagation (plain white line). The intensity of the condensate emission is indicated in false colors.

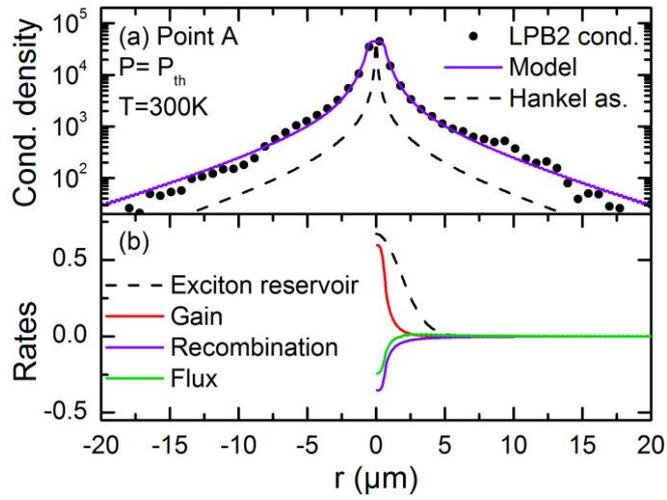

**Figure 8:** (a) Experimental y-axis profiles of the condensate (dots) above the polariton laser threshold ($P = 1.3\, P_{th}$, LPB2, T=80K), and GPE simulation (plain line) and asymptotic Hankel function (dashed line); the blueshift of the condensate is $11\, meV$, as in the corresponding experiment (figure 5) and a reservoir density at the center is $7\, 10^5\, cm^{-2}$. The polariton decay rate $\gamma_{pol} = 0.4\, meV$. (b) Radial dependence of the simulated rates for gain (stimulated relaxation), polariton recombination and polariton flux.

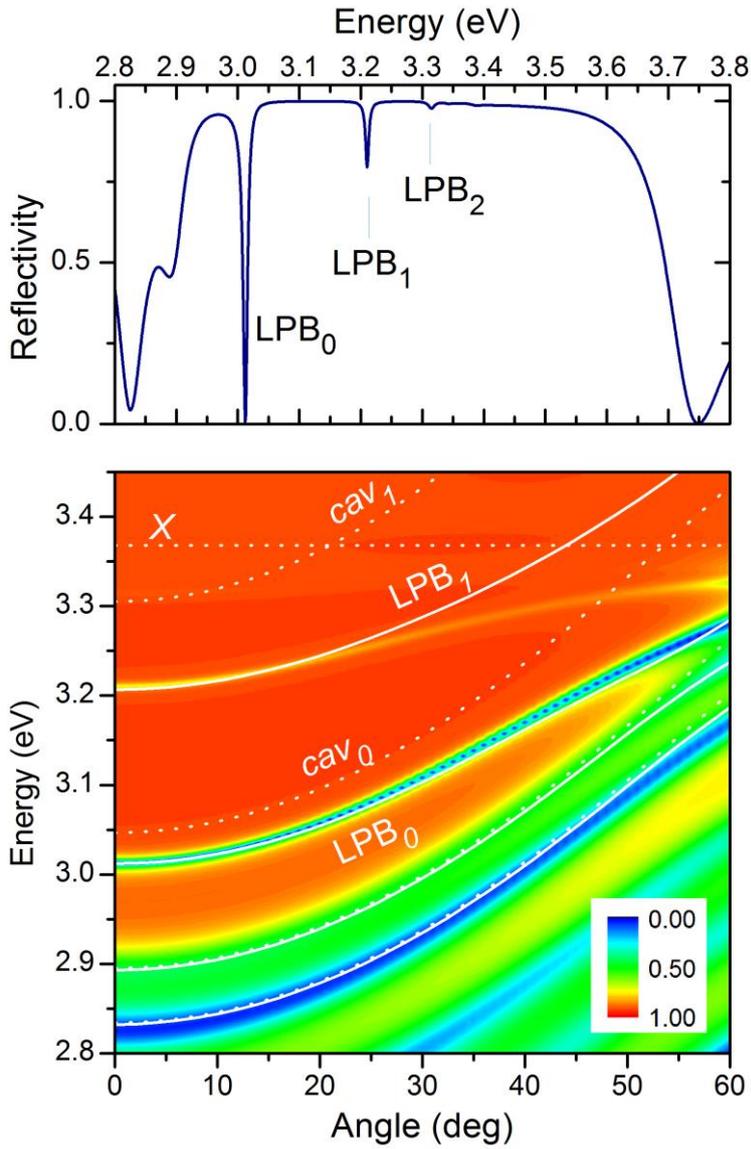

**Figure A1:** (a) Transfer-matrix simulation of the reflectivity of the microcavity at normal incidence (TM polarization) for a cavity thickness $L = 890\ nm$, and a temperature $T = 80K$. (b) Angular dependence of the reflectivity (false colors). The bare exciton and photon modes and the dispersion of the polariton branches deduced from the coupled oscillator model are indicated as dashed and plain lines respectively. The parameters are summarized in table 1.

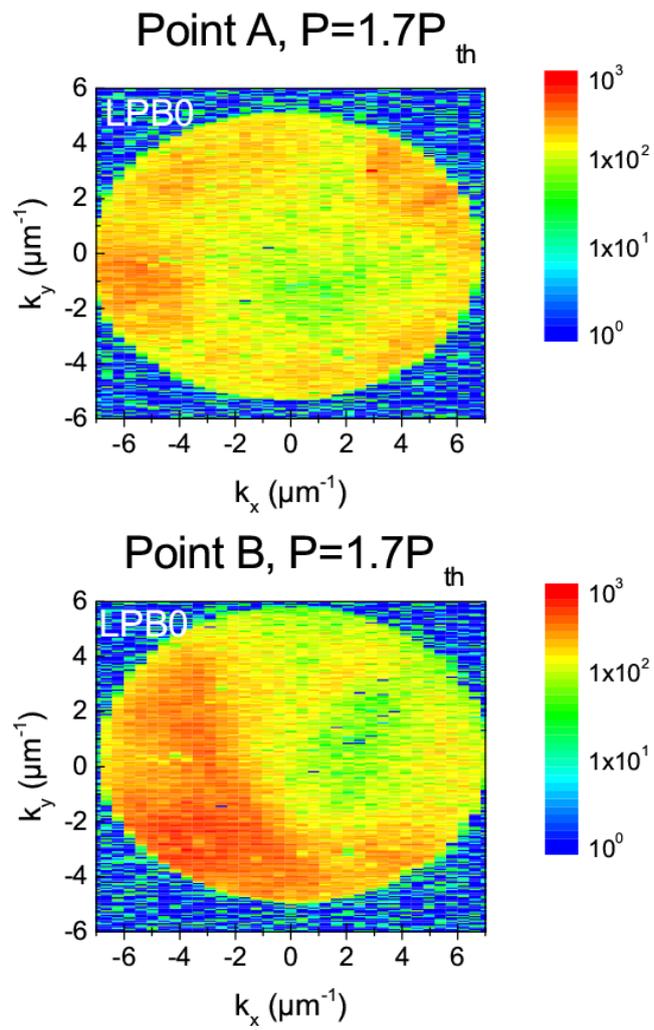

**Figure A2:** Far-field images of the uncondensed LPB0 polaritons at points A and B, above threshold (T=300K, logarithmic color scales).

**References**

[1] B. Nelsen, G. Liu, M. Steger, D. W. Snoke, R. Balili, K. West, and L. Pfeiffer, Phys. Rev. X **3**, 041015 (2013).

[2] B. Sermage, G. Malpuech, A. V. Kavokin, and V. Thierry-Mieg, Phys. Rev. B **64**, 081303 (2001).

[3] M. Steger, G. Liu, B. Nelsen, C. Gautham, D. W. Snoke, R. Balili, L. Pfeiffer, and K. West, Phys. Rev. B **88**, 235314 (2013).

[4] E. Wertz, A. Amo, D. D. Solnyshkov, L. Ferrier, T. C. H. Liew, D. Sanvitto, P. Senellart, I. Sagnes, A. Lemaître, A. V. Kavokin, G. Malpuech, and J. Bloch, Phys Rev Lett **109**, 216404 (2012).

[5] T. Gao, P. S. Eldridge, T. C. H. Liew, S. I. Tsintzos, G. Stavrinidis, G. Deligeorgis, Z. Hatzopoulos, and P. G. Savvidis, Phys Rev B **85**, 235102 (2012).

[6] H. S. Nguyen, D. Vishnevsky, C. Sturm, D. Tanese, D. Solnyshkov, E. Galopin, A. Lemaître, I. Sagnes, A. Amo, G. Malpuech, and J. Bloch, Phys Rev Lett **110**, 236601 (2013).

[7] D. D. Solnyshkov, H. Terças, and G. Malpuech, Appl. Phys. Lett. **105**, 231102 (2014).

[8] K. G. Lagoudakis, M. Wouters, M. Richard, A. Baas, I. Carusotto, R. André, L. S. Dang, and B. Deveaud-Plédran, Nat. Phys. **4**, 706 (2008).

[9] D. Sanvitto, F. M. Marchetti, M. H. Szymanska, G. Tosi, M. Baudisch, F. P. Laussy, D. N. Krizhanovskii, M. S. Skolnick, L. Marrucci, A. Lemaitre, J. Bloch, C. Tejedor, and L. Vina, Nat Phys **6**, 527 (2010).

[10] A. Amo, S. Pigeon, D. Sanvitto, V. G. Sala, R. Hivet, I. Carusotto, F. Pisanello, G. Leménager, R. Houdré, E. Giacobino, C. Ciuti, and A. Bramati, Science **332**, 1167 (2011).

[11] R. Hivet, H. Flayac, D. D. Solnyshkov, D. Tanese, T. Boulier, D. Andreoli, E. Giacobino, J. Bloch, A. Bramati, G. Malpuech, and A. Amo, Nat. Phys. **8**, 724 (2012).

[12] M. Sich, K. D. N., S. M. S., G. A. V., R. Hartley, S. D V., C.-M. E. A., K. Biermann, R. Hey, and P. V. Santos, Nat Photon **6**, 50 (2012).

[13] P. M. Walker, L. Tinkler, D. V. Skryabin, A. Yulin, B. Royall, I. Farrer, D. A. Ritchie, M. S. Skolnick, and D. N. Krizhanovskii, Nat. Commun. **6**, 8317 (2015).

[14] I. Carusotto and C. Ciuti, Rev Mod Phys **85**, 299 (2013).

[15] T. Guillet, M. Mexis, J. Levrat, G. Rossbach, C. Brimont, T. Bretagnon, B. Gil, R. Butté, N. Grandjean, L. Orosz, F. Reveret, J. Leymarie, J. Zúñiga-Pérez, M. Leroux, F. Semond, and S. Bouchoule, Appl Phys Lett **99**, 161104 (2011).

[16] H. Franke, C. Sturm, R. Schmidt-Grund, Gerald Wagner, and M. Grundmann, New J. Phys. **14**, 013037 (2012).

[17] L. Orosz, F. Réveret, F. Médard, P. Disseix, J. Leymarie, M. Mihailovic, D. Solnyshkov, G. Malpuech, J. Zúñiga-Pérez, F. Semond, M. Leroux, S. Bouchoule, X. Lafosse, M. Mexis, C. Brimont, and T. Guillet, Phys Rev B **85**, 121201 (2012).

[18] W. Xie, H. Dong, S. Zhang, L. Sun, W. Zhou, Y. Ling, J. Lu, X. Shen, and Z. Chen, Phys Rev Lett **108**, 166401 (2012).

[19] T.-C. Lu, Y.-Y. Lai, Y.-P. Lan, S.-W. Huang, J.-R. Chen, Y.-C. Wu, W.-F. Hsieh, and H. Deng, Opt Express **20**, 5530 (2012).

[20] F. Li, L. Orosz, O. Kamoun, S. Bouchoule, C. Brimont, P. Disseix, T. Guillet, X. Lafosse, M. Leroux, J. Leymarie, M. Mexis, M. Mihailovic, G. Patriarche, F. Réveret, D. Solnyshkov, J. Zúñiga-Pérez, and G. Malpuech, Phys Rev Lett **110**, 196406 (2013).

[21] A. Trichet, E. Durupt, F. Médard, S. Datta, A. Minguzzi, and M. Richard, Phys. Rev. B **88**, 121407 (2013).

[22] D. Xu, W. Xie, W. Liu, J. Wang, L. Zhang, Y. Wang, S. Zhang, L. Sun, X. Shen, and Z. Chen, Appl. Phys. Lett. **104**, 082101 (2014).
[23] J. Wang, W. Xie, L. Zhang, Y. Wang, J. Gu, T. Hu, L. Wu, and Z. Chen, Solid State Commun. **211**, 16 (2015).
[24] K. S. Daskalakis, S. A. Maier, and S. Kéna-Cohen, Phys. Rev. Lett. **115**, 035301 (2015).
[25] M. Wouters and I. Carusotto, Phys Rev Lett **99**, 140402 (2007).
[26] M. Wouters, I. Carusotto, and C. Ciuti, Phys Rev B **77**, 115340 (2008).
[27] F. Li, L. Orosz, O. Kamoun, S. Bouchoule, C. Brimont, P. Disseix, T. Guillet, X. Lafosse, M. Leroux, J. Leymarie, G. Malpuech, M. Mexis, M. Mihailovic, G. Patriarche, F. Reveret, D. Solnyshkov, and J. Zuniga-Perez, Appl Phys Lett **102**, 191118 (2013).
[28] C. Brimont, T. Guillet, S. Rousset, D. Néel, X. Checoury, S. David, P. Boucaud, D. Sam-Giao, B. Gayral, M. J. Rashid, and F. Semond, Opt Lett **38**, 5059 (2013).
[29] M. Aßmann, F. Veit, M. Bayer, A. Löffler, S. Höfling, M. Kamp, and A. Forchel, Phys. Rev. B **85**, 155320 (2012).
[30] M. Thunert, A. Janot, H. Franke, C. Sturm, T. Michalsky, M. D. Martin, L. Viña, B. Rosenow, M. Grundmann, and R. Schmidt-Grund, Arxiv 1412.8667 (2014).
[31] H. Haug and S. Koch, Phys. Status Solidi B **82**, 531 (1977).
[32] A. Kavokin, J. Baumberg, G. Malpuech, and F. Laussy, *Microcavities* (Oxford Science Publications, 2006).
[33] G. Nardin, K. G. Lagoudakis, M. Wouters, M. Richard, A. Baas, R. André, L. S. Dang, B. Pietka, and B. Deveaud-Plédran, Phys Rev Lett **103**, 256402 (2009).
[34] I. Iorsh, M. Glauser, G. Rossbach, J. Levrat, M. Cobet, R. Butté, N. Grandjean, M. A. Kaliteevski, R. A. Abram, and A. V. Kavokin, Phys. Rev. B **86**, 125308 (2012).
[35] M. Amthor, T. C. H. Liew, C. Metzger, S. Brodbeck, L. Worschech, M. Kamp, I. A. Shelykh, A. V. Kavokin, C. Schneider, and S. Höfling, Phys. Rev. B **91**, 081404 (2015).
[36] D. D. Solnyshkov, H. Terças, K. Dini, and G. Malpuech, Phys. Rev. A **89**, 033626 (2014).
[37] C. P. Dietrich, R. Johne, T. Michalsky, C. Sturm, P. Eastham, H. Franke, M. Lange, M. Grundmann, and R. Schmidt-Grund, Phys. Rev. B **91**, 041202 (2015).
[38] T. Guillet, C. Brimont, P. Valvin, B. Gil, T. Bretagnon, F. Médard, M. Mihailovic, J. Zúñiga-Pérez, M. Leroux, F. Semond, and S. Bouchoule, Phys Stat Sol C **9**, 1225 (2012).